\documentclass[prl,aps,twocolumn,showpacs,preprintnumbers,amsmath,amssymb]{revtex4}

\usepackage{graphicx}
\usepackage{dcolumn}
\usepackage{color}

\begin{document}

\def\cred{\color{red}}
\def\cblue{\color{blue}}

\title{
Topological properties and the dynamical crossover from mixed-valence
to Kondo-lattice behavior in golden phase of SmS
}

\author{Chang-Jong Kang}\email{rkdc1234@postech.ac.kr}
\affiliation{Department of Physics, Pohang University of Science and Technology (POSTECH), Pohang, 790-784, Korea }

\author{Hong Chul Choi}
\affiliation{Department of Physics, Pohang University of Science and Technology (POSTECH), Pohang, 790-784, Korea }
\author{Kyoo Kim}
\affiliation{Department of Physics, Pohang University of Science and Technology (POSTECH), Pohang, 790-784, Korea }

\author{B. I. Min}\email{bimin@postech.ac.kr}
\affiliation{Department of Physics, Pohang University of Science and Technology (POSTECH), Pohang, 790-784, Korea }
\date{\today}

\begin{abstract}
We have investigated temperature-dependent behaviors of
electronic structure and resistivity in a mixed-valent golden phase of SmS,
based on the dynamical mean-field theory band structure calculations.
Upon cooling, the coherent Sm $4f$ bands are formed to produce
the hybridization-induced pseudogap near the Fermi level,
and accordingly the topology of Fermi surface
is changed to exhibit a Lifshitz-like transition.
The surface states emerging in the bulk gap region are found to be
not topologically protected states but just typical Rashba spin-polarized states,
indicating that SmS is not a topological Kondo semimetal.
From the analysis of anomalous resistivity behavior in SmS,
we have identified universal energy scales,
which characterize the Kondo/mixed-valent semimetallic systems.
\end{abstract}

\pacs{71.27.+a, 71.18.+y, 75.30.Mb}


\maketitle

Kondo and mixed-valent physics in strongly-correlated $4f$ electron systems
have been subject of longstanding controversy.
It includes various interesting phenomena such as
$p$-wave superconductivity and non-Fermi-liquid behavior
in heavy-fermion systems \cite{Stewart84,Coleman01,Gegenwart08},
and recently proposed topological Kondo insulator behavior in a typical mixed-valent insulator
SmB$_{6}$ \cite{Dzero10,Takimoto11,Dzero12,Kang13,Xu13,Neupane13,Kim14,Kim13,
Kim13-2,Jonathan13,Chmin14}.
The topological Kondo insulator of our present interest has attracted
a great deal of recent attention.
However, the realization of the topological properties in SmB$_{6}$ is
still under debate.
The immediate question was addressed whether a similar mixed-valent system,
SmS, also has the topological properties or not.
Indeed, the mixed-valent golden phase of SmS (g-SmS) was reported
to be a topological Kondo semimetal \cite{Li14}.

\begin{figure}[b]
\includegraphics[width=8.5 cm]{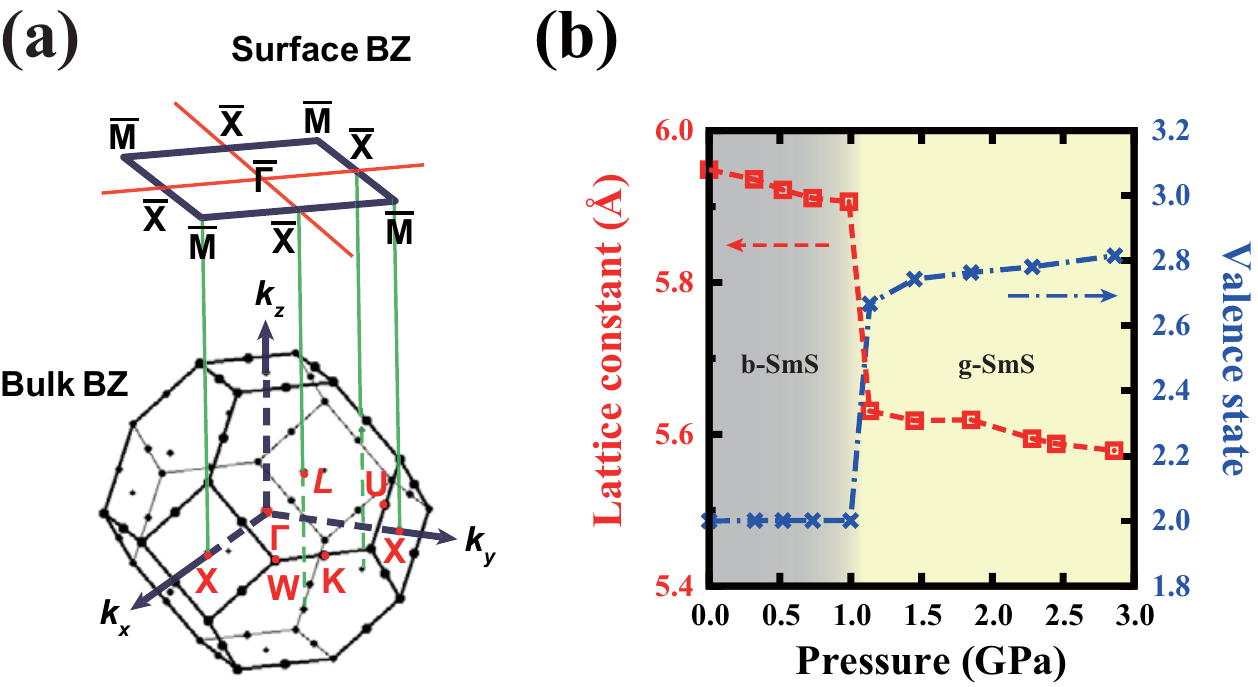}
\caption{(Color Online) Isostructural phase transition in SmS.
(a) Bulk and surface Brillouin zones (BZs) of fcc SmS.
(b) The pressure dependence of the lattice constant and the Sm valence state of SmS
(Data taken from Ref.~\cite{Deen05}).
}
\label{struct}
\end{figure}

SmS has been studied for last four decades \cite{Varma76,Villars91,Campagna74,Coey76,
Maple71,Jayaraman70,Deen05,Shapiro75,Martin79,Barla04,Imura09,Ito02,Mizuno08,Matsubayashi07,
Flouquet04,Wachter94,Antonov02,Svane05,Lapierre81,Konczykowski81,Konczykowski85,Imura09-2},
but there remain several issues still unresolved.
SmS crystallizes in a face-centered cubic (fcc) structure of rock-salt (NaCl-type) type.
At the ambient pressure, SmS has a so-called black phase, which is a semiconductor
with indirect and direct band gaps of 90 meV and 0.4 eV, respectively \cite{Mizuno08}.
In the black phase of SmS (b-SmS), the valence state of Sm is divalent (2+),
and so the system is nonmagnetic.
Under high pressure above 6.5 kbar, SmS undergoes a first-order isostructural
phase transition from b-SmS to g-SmS, in which Sm ions are mixed-valent
with the average valency of 2.6+ $\sim$ 2.8+ \cite{Coey76,Maple71}.
This isostructural transition is accompanied by the volume collapse
as much as 15\% \cite{Maple71}, as shown in Fig.~\ref{struct}(b).

\begin{figure*}[t]
\includegraphics[width=17 cm]{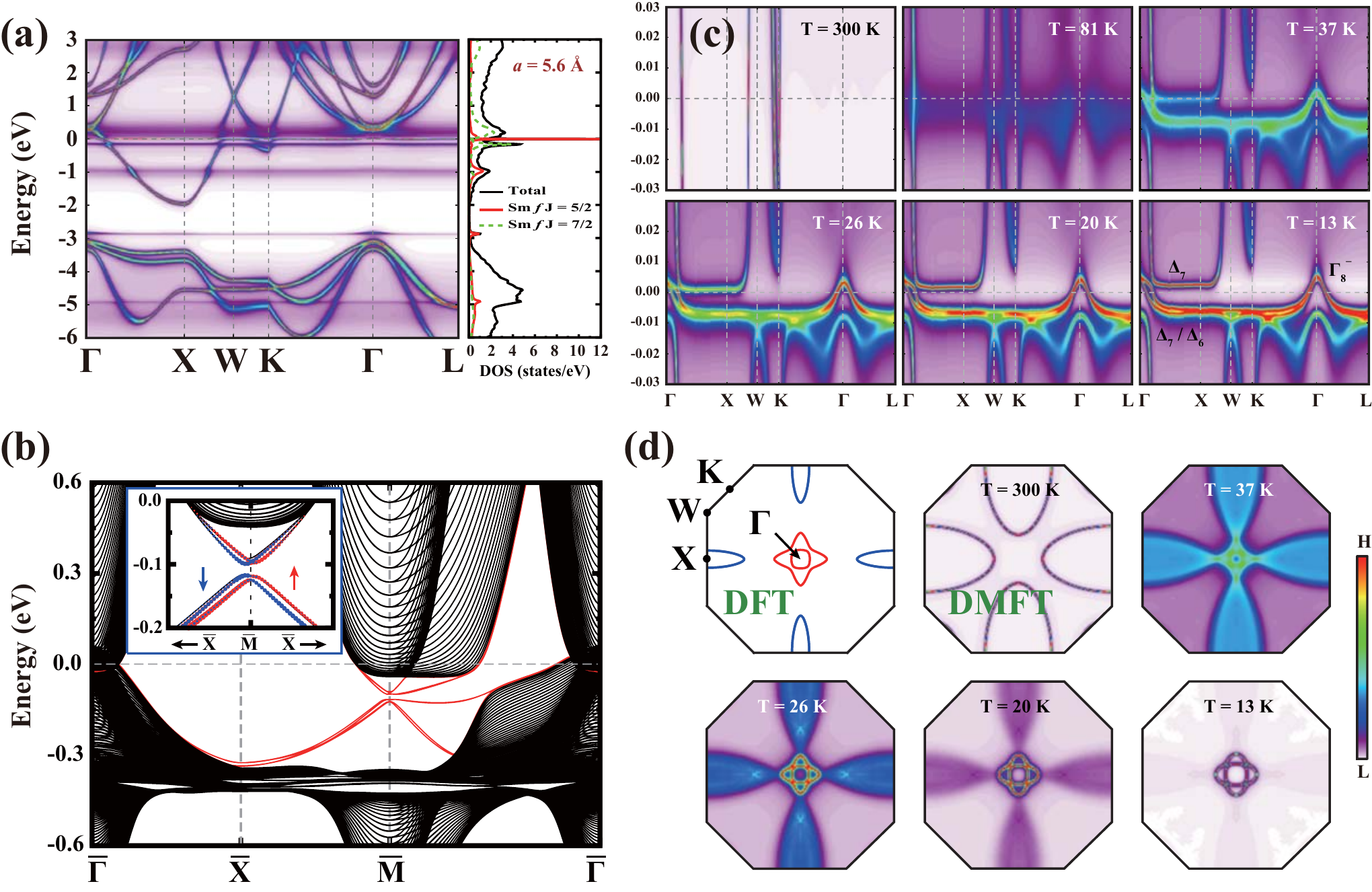}
\caption{(Color Online) Electronic structure of g-SmS ($a$ = 5.6 $\AA$).
(a) The DMFT band structure at $T = 13$ K.
(b) Surface band structure obtained by the model slab calculation
based on the DFT band structure.
The surface states in the inset manifests a typical Rashba splitting
with a tiny gap, and so they are not topologically protected states.
(c) $T$-dependent band structures in the DMFT scheme.
Upon cooling, the coherent Sm $4f$ band emerges near $E_{F}$,
and the hybridization gap appears below $T= 26$ K.
(d) $T$-dependent FS evolution in the DMFT scheme,
which is compared with the DFT FS
with 10 times enhanced SOC of Sm 4$f$-electron (top-left) \cite{Supp}.
Upon cooling, $X$-centered $d$-band pockets are reduced,
while $f$-band pockets emerge at $\Gamma$.
}
\label{golden-DMFT}
\end{figure*}

The energy gap decreases monotonically with increasing the pressure.
It is controversial whether g-SmS has a real gap or a pseudo-gap \cite{Matsubayashi07,Wachter94}.
The resistivity behavior of g-SmS is quite anomalous
in the sense that the overall behavior is Kondo-lattice like but
it exhibits a couple of abnormal hump structures
\cite{Lapierre81,Konczykowski81,Konczykowski85,Imura09,Imura09-2}.
Applying the pressure further, g-SmS has a magnetic instability at about 19.5 kbar
with the antiferromagnetic order \cite{Shapiro75,Barla04,Imura09}.
Above 19.5 kbar, the resistivity shows a metallic behavior
and the valence state of Sm increases toward 3+ \cite{Imura09}.

There have been several reports on the density-functional theory (DFT)-based
band structure study of SmS \cite{Antonov02,Svane05,Li14}.
But even the ground-state insulating nature of b-SmS is not
properly described by the DFT-based schemes,
and so the advanced methods like the dynamical mean-field theory (DMFT) should be employed
to investigate the topological properties in g-SmS
that has the strongly-correlated 4$f$ electrons.
Here we have investigated electronic structures of SmS,
based on the DMFT scheme.
First, we have shown that the electronic properties of b-SmS are described properly only
by the DMFT scheme.
Then we have examined the $T$-dependent electronic structure evolution
in g-SmS.
Upon cooling, the $4f$ states form the coherent $4f$ bands with a pseudo-gap feature
near the Fermi level ($E_F$), and
accordingly the topology of the Fermi surface (FS) is changed,
which is reflected well in the anomalous resistivity behavior in g-SmS.
We have demonstrated that the surface states realized in g-SmS
are not the topological states, but are just the typical Rashba states.

We have employed the all-electron FLAPW band method implemented in Wien2k \cite{Wien2k}.
We have checked that both the DFT and the DFT+$U$ (on-site Coulomb $U$) schemes
can not describe the ground state insulating electronic structure of SmS
properly (see the supplement \cite{Supp}).
Therefore, we have employed the combined DFT and DMFT (DFT+DMFT) approach
implemented in Wien2k, which has successfully reproduced
many aspects of the strongly-correlated electron systems \cite{Kotliar06,Haule10}.
We used projectors in the large window of 10 eV, and
the on-site Coulomb and exchange energies of $U$ = 6.1 eV, $J$ = 0.8355 eV
were adopted to fit in X-ray photoemission spectroscopy (XPS)
data (see Fig. S2 for the DMFT band structure of b-SmS in the supplement) \cite{Campagna74}.
To solve the impurity problem, the non-crossing approximation (NCA)
is used \cite{Jonathan13,Kim14}.
To verify the justification of the NCA scheme,
we have also used the continuous time quantum Monte Carlo scheme
for b-SmS and checked that two schemes give the same result.

\begin{figure}[t]
\includegraphics[width=8.5 cm]{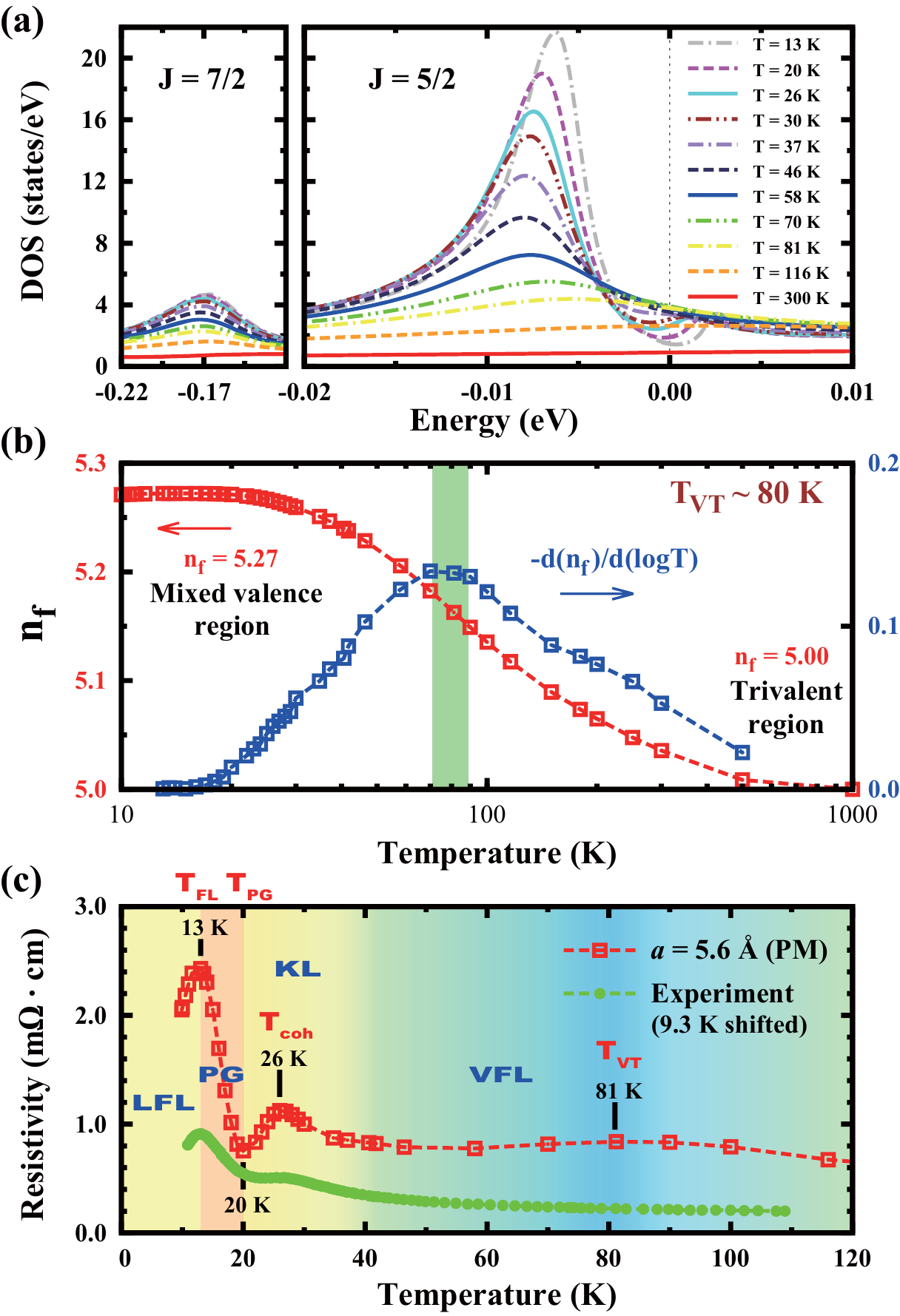}
\caption{(Color Online) $T$-dependent behaviors in g-SmS ($a$ = 5.6 $\AA$).
(a) $T$-dependent DOS variation in the DMFT scheme.
Upon cooling, the $J = 5/2$ DOS behaves like a Kondo resonance
with the pseudo-gap like dip feature near $E_{F}$.
The $J = 7/2$ DOS peak near $-0.17$ eV is also shown.
(b) The number of occupied $f$-electrons ($n_{f}$)
as a function of the logarithmic $T$ scale.
Upon cooling, $n_{f}$ increases monotonically from trivalent state to
the mixed-valent state.
The effective valence-transition (VT) occurs at $T \simeq 80$ K,
as indicated by the peak in $-d(n_{f})/d(\log T)$.
(c) Resistivity vs. $T$ obtained by the DMFT scheme.
The experimental data are taken from Ref.~\cite{Imura09-2},
which are shifted by 9.3 K toward higher $T$.
Upon cooling, SmS undergoes several crossover transitions from
valence-fluctuation liquid (VFL),
Kondo-liquid (KL), pseudogap (PG), to Landau Fermi-liquid (LFL).
}
\label{nf-T}
\end{figure}

The DMFT band structure and DOS of g-SmS ($a$ = 5.6 $\AA$) at $T = 13$ K
are shown in Fig. \ref{golden-DMFT}(a).
Notable is the flat Sm $4f$ bands near $E_F$,
which yield sharp Kondo resonance-like peaks in the DOS (see Fig. \ref{nf-T}(a)).
The $4f$-band in the vicinity of $E_{F}$ is mainly of $J = 5/2$ character ($4f_{5/2}$),
while $4f_{7/2}$ bands are located at about $\pm$0.17 eV from $E_{F}$.
These coherent Sm $4f$ bands hybridize strongly with Sm $5d$ band
to produce the hybridization gap near $E_F$ (see Fig.~\ref{golden-DMFT}(c)).
However, the $4f_{5/2}$ ($\Gamma_{8}^{-}$) band at $\Gamma$ is above $E_F$,
and so the band structure exhibits the metallic nature
having the mixed-valent state of Sm$^{2.73+}$.
The $\Gamma_{8}^{-}$ band is dispersive
with the band width of about 0.03 eV, as in SmB$_{6}$ \cite{Kang13}.
The DMFT band structure near $E_{F}$ is analogous to the band structure
obtained by the DFT+SOC (SOC: spin-orbit coupling),
as shown in the supplement \cite{Supp},
but the band width of the former is nearly ten times smaller than that of
the latter, which is also similar to the case in SmB$_{6}$ \cite{Kang13,Kim14}.

Even though g-SmS has metallic nature, the band inversion occurs at $X$,
and so it is tempting to anticipate the topologically protected surface states
in SmS \cite{Li14}.
To check the topological property, we have examined the surface
band structure in Fig. \ref{golden-DMFT}(b), which was obtained by
the model slab calculation based on the DFT bulk band structure \cite{Slab}.
The surface states, however, have a tiny gap instead of a Dirac cone
and the momentum-dependent splitting of spin states, which
reveals that they are not topologically protected states
but just Rashba spin-polarized surface states.
This result is a contrast to that of Li \emph{et. al.}~\cite{Li14},
who argued that g-SmS is to be a topological semimetal.
The charge gap protection is essential to have the topological nature.
In this respect, La-doped SmS can be a better candidate for a topological Kondo insulator,
because Sm$_{0.75}$La$_{0.25}$S under pressure was reported to be an excitonic insulator
with an energy gap of 1 meV \cite{Wachter95}.

Figure \ref{golden-DMFT}(c) shows the $T$-dependent band structures
of g-SmS ($a$ = 5.6 $\AA$).
At $T = 300$ K, only the Sm $d$-band is seen to cut $E_{F}$.
Upon cooling, Sm $4f$ spectra start to emerge near $E_{F}$ at $T \simeq 80$ K,
and $4f_{5/2}$ bands are observed to be formed below $T \simeq 40$ K.
At $T = 26$ K, the $4f_{5/2}$ hole bands near $\Gamma$ ($\Gamma_{8}^{-}$ quartet)
forms the coherent band, and the hybridization gap between Sm $5d$ ($t_{2g}$)
and $4f_{5/2}$ ($\Delta_{7}$) bands begins to appear at the crossing
point near $\Gamma$.
The coherence temperature $T_{coh}$ is usually defined
in metallic Kondo lattices as an onset $T$,
at which the hybridization gap appears \cite{Choi12}.
Note that $T_{coh}$ is equivalent to $T^*$
that is introduced in the two-fluid model for the Kondo lattice \cite{Yang08}.
Hence $T = 26$ K is considered to be $T_{coh}$ of g-SmS.
In the Kondo lattice systems, $T_{coh}$ can often be identified from the peak
position of the resistivity, as will be discussed in Fig. \ref{nf-T}(c).
With decreasing $T$ further, the separation between the upper $4f_{5/2}$ flat band
($\Delta_7$ doublet) along $\Gamma-X$ and the lower $4f_{5/2}$ flat band
($\Delta_7$/$\Delta_6$ quartet) becomes enhanced,
and eventually the former is shifted up above $E_{F}$ at $T = 20$ K.
Below $T \simeq 13$ K, it is seen that the $4f_{5/2}$ band becomes completely
coherent over the whole Brillouin zone.
In this case, the $\Gamma_{8}^{-}$ hole band is still above $E_{F}$,
and so SmS at very low $T$ would exhibit metallic nature (see Fig. \ref{nf-T}(c)).

Figure \ref{golden-DMFT}(d) presents the $T$-dependent FS evolution in g-SmS.
Also presented is the DFT-FS, for comparison,
which has $X$-centered electron ellipses and
$\Gamma$-centered hole pockets.
At $T = 300$ K, the DMFT-FS arises solely from $d$-band,
which produces the $X$-centered ellipses.
Upon cooling, the ellipses become reduced more and more,
and so almost disappear at $T = 13$ K.
Their spectral weights, however, become enhanced due to the contribution from
the hybridized $4f_{5/2}$ band so as to have maximum intensity at $T \simeq 37$ K.
On the other hand, $\Gamma$-centered hole pockets
originating from $4f_{5/2}$ ($\Gamma_{8}^{-}$) band begin to appear at $T \simeq 37$ K,
and are clearly manifested below $T_{coh} = 26$ K.
Therefore, the topology of the FS is to be changed twice,
at $T \simeq 37$ K and $T \simeq 13$ K.
The former is due to the emergence of the coherent $4f$ band formation at $\Gamma$,
while the latter is due to the separation of the upper $4f_{5/2}$
flat band ($\Delta_{7}$ doublet) from $E_F$.

This behavior is reminiscent of the Lifshitz transition
that is a typical continuous quantum phase transition
characterized by the topological change of the FS \cite{Lifshitz60}.
The feature in SmS is quite interesting because
a Lifshitz-like transition occurs with the variation of $T$.
Such a Lifshitz-like transition in g-SmS could be explored
by ARPES and de Hass-van Alphen experiments.
The change of the FS topology is well reflected
in the $T$-dependent resistivity behavior in Fig.~\ref{nf-T}(c).

Upon cooling, the $4f_{5/2}$ DOS peak at $\sim -6$ meV
becomes sharper and sharper,
as shown in Fig.~\ref{nf-T}(a),
resembling the $T$-dependent evolution of Kondo resonance.
Noteworthy is the pseudo-gap feature near $E_{F}$, which arises from the
$f$-$d$ band hybridization below $T \simeq  30$ K,
which is close to $T_{coh}=26$ K.
It is also seen that, upon cooling, the DOS at $E_{F}$ increases first
and then decreases below $T \simeq 80$ K \cite{Supp}.
The $4f_{7/2}$ DOS near $-0.17$ eV,
which corresponds to the spin-orbit split side-band,
behaves similarly to the $4f_{5/2}$ DOS upon cooling.

The number of $f$-electrons ($n_{f}$) in g-SmS versus $T$
is presented in Fig.~\ref{nf-T}(b).
Upon cooling, $n_{f}$ increases monotonically from
$n_{f} = 5$ (trivalent state) at high $T$ to $n_{f} = 5.27$
(mixed-valence state) at low $T$.
The increasing trend of $n_{f}$ upon cooling has also been
observed in SmB$_6$ \cite{Jonathan13,Mizumaki09}.
It is worthwhile to observe that $n_{f}$ curve has an inflection point
at $T \simeq 80$ K, which indicates that the effective valence-transition (VT)
occurs at $T \simeq 80$ K.

Figure~\ref{nf-T}(c) shows the resistivity versus $T$ of g-SmS
evaluated in the DMFT scheme \cite{Choi12}.
The overall behavior is that the resistivity increases upon cooling,
but decreases below $T = 13$ K,
as in metallic Kondo lattice systems.
However, the detailed $T$-dependent behavior is quite anomalous.
The calculated resistivity has a broad maximum at $T \simeq 80$ K,
and starts to increase again at $T \simeq 40$ K to produce
hump and dip structure at $T \simeq 26$ K and $T \simeq 20$ K, respectively.
Upon further cooling, it exhibits another maximum at $T = 13$ K.
Quite similar behavior is indeed observed in the measured resistivity
of g-SmS \cite{Lapierre81,Konczykowski81,Konczykowski85,Imura09-2},
as shown in Fig.~\ref{nf-T}(c),
even though the feature at $T \simeq 80$ K is not so obvious.

The anomalous behavior of the resistivity is
closely correlated with the $T$-dependent evolution of electronic structure
in Fig.~\ref{golden-DMFT}(c)-(d).
Namely, the broad maximum at $T \simeq 80$ K is considered to arise from
the dynamical valence-fluctuation (VF),
as manifested by the effective VT in Fig.~\ref{nf-T}(b)
and the incoherent Sm $4f$ spectra in Fig.~\ref{golden-DMFT}(c)
at $T \simeq 80$ K.
Intriguingly, g-SmS has a minimum volume at $T \simeq 80$ K \cite{Iwasa05},
which signifies the close connection of a broad maximum in the resistivity
with the mixed-valent nature of SmS.
Hence we assign $T\simeq 80$ K as $T_{VT}$.
Interestingly, SmB$_{6}$ also exhibits a similar broad maximum
feature in the resistivity \cite{Kebede96}
and a minimum volume \cite{Trounov93} near $T \simeq 150$ K.
The resistivity increases below $T \simeq 40$ K, at which the coherent $4f_{5/2}$-band
starts to emerge in Fig.~\ref{golden-DMFT}(c).
The resistivity hump at $T \simeq 26$ K indicates that $T_{coh}$ is around 26 K.
In conventional metallic Kondo lattices, the resistivity decreases monotonically
to zero below $T_{coh}$.
The behavior near $T_{coh}$ was explained in the two-fluid model by
a crossover from the Kondo spin-liquid (KSL) to Kondo Fermi-liquid (KFL) \cite{Yang08}.
We use a term of Kondo-liquid (KL) in Fig.~\ref{nf-T}(c)
to comprise both KSL and KFL phases.

Thus g-SmS exhibits a crossover from mixed-valence to Kondo lattice
behavior upon cooling \cite{Coleman07,Kumar11}.
The charge-fluctuation that starts to be effective at $T\simeq 80$ K
becomes almost frozen at $T \simeq 20$ K, as shown in Fig.~\ref{nf-T}(b).
Then, the spin-fluctuation becomes dominating near $T \simeq 25$ K,
so as to activate the Kondo screening.
In fact, the effective hybridization obtained in the DMFT becomes
the largest near $T=25$ K \cite{Supp}.

Meanwhile, $T \simeq 20$ K of the dip structure coincides with $T$, at which
the upper $4f_{5/2}$ flat band ($\Delta_{7}$ doublet) along $\Gamma-X$ is detached from $E_{F}$.
Thus the resistivity up-turn occurs due to the apparent pseudo-gap (PG) feature,
and so we assign $T = 20$ K as $T_{PG}$.
Actually, this kind of hump and dip structure in the resistivity has been observed in several
Ce compounds of Kondo-insulator type, such as CeNiSn and CeRhSb \cite{Takabatake90,Malik91}.
The third resistivity drop at $T = 13$ K occurs due to
the complete formation of the coherent $4f_{5/2}$ band
over the whole Brillouin zone.
Below $T = 13$ K, the imaginary part of self-energy almost vanishes \cite{Supp},
whereby the crossover from the KL to the Landau Fermi-liquid (LFL)
takes place \cite{Yang08}. So we assign $T = 13$ K as $T_{FL}$.
Therefore, SmS undergoes several crossover transitions upon cooling, from VF liquid (VFL),
KL, PG to LFL, as shown in Fig.~\ref{nf-T}(c).
The energy scales $T_{VT}$, $T_{coh}$, $T_{PG}$, and $T_{FL}$,
which are identified for SmS, are considered to be universal
to characterize Kondo/mixed-valent semimetallic systems.

Finally, it should be pointed out that the resistivity behavior in SmS is
quite different from that in SmB$_6$ that is a candidate of topological Kondo insulators.
g-SmS is not a Kondo insulator but close to a Kondo semimetal,
so that the conventional metallic resistivity behavior is expected
to be realized at very low $T$.

In conclusion,
in g-SmS, the coherent Sm $4f$ bands are formed upon cooling
to produce the hybridization-induced pseudogap feature near $E_F$,
which is accompanied by a Lifshitz-like topological transition in the FS.
g-SmS is found to be not a topological Kondo semimetal in view of that
the in-gap surface states are not topological states but
are typical spin-polarized Rashba states.
From the analysis of $T$-dependent resistivity of g-SmS,
we have identified characteristic multiple energy scales,
which are expected to govern
the physics of Kondo/mixed-valent semimetallic systems universally.

Helpful discussions with Junwon Kim, J. H. Shim, B. H. Kim, Ki-Seok Kim,
J. -S. Kang, J. D. Denlinger, J. W. Allen, and K. Sun are greatly appreciated.
This work was supported by the NRF (No. 2009-0079947 and No. 2011-0025237),
POSTECH BK21Plus Physics Division and BSRI grant,
and the KISTI supercomputing center (No. KSC-2013-C3-010).

\newpage

\setcounter{table}{0}
\setcounter{figure}{0}
\onecolumngrid

\begin{center}
{\bf \Large
{\it Supplemental Material:}\\
Topological properties and the dynamical crossover from mixed-valence
to Kondo-lattice behavior in golden phase of SmS
}
\end{center}

\author{Chang-Jong Kang}
\affiliation{Department of Physics, Pohang University of Science and Technology (POSTECH), Pohang, 790-784, Korea }

\author{Hong Chul Choi}
\affiliation{Department of Physics, Pohang University of Science and Technology (POSTECH), Pohang, 790-784, Korea }
\author{Kyoo Kim}
\affiliation{Department of Physics, Pohang University of Science and Technology (POSTECH), Pohang, 790-784, Korea }

\author{B. I. Min}
\affiliation{Department of Physics, Pohang University of Science and Technology (POSTECH), Pohang, 790-784, Korea }

\renewcommand{\thefigure}{S\arabic{figure}}
\renewcommand{\thetable}{S\arabic{table}}

\section{A black phase of SmS (b-SmS)}
\subsection{DFT band structure}

For the DFT calculations, we have employed the all-electron FLAPW band method
implemented in Wien2k \cite{Wien2k}.
We used 17 $\times$ 17 $\times$ 17 {\bf $k$}-point mesh in the full Brillouin zone.
The muffin-tin radii $R_{MT}$ were set to 2.50 and 2.30 a.u. for Sm and S, respectively,
and the product of $R_{MT}$ and the maximum reciprocal lattice
vector $K_{max}$ was chosen as $R_{MT}\cdot K_{max} = 8$.
We used the maximum $L$ value of 10 for the waves inside the atomic spheres
and the largest reciprocal lattice vector $G_{max}$ of 12 in the charge Fourier expansion.

\begin{figure*}[t]
\includegraphics[width=18 cm]{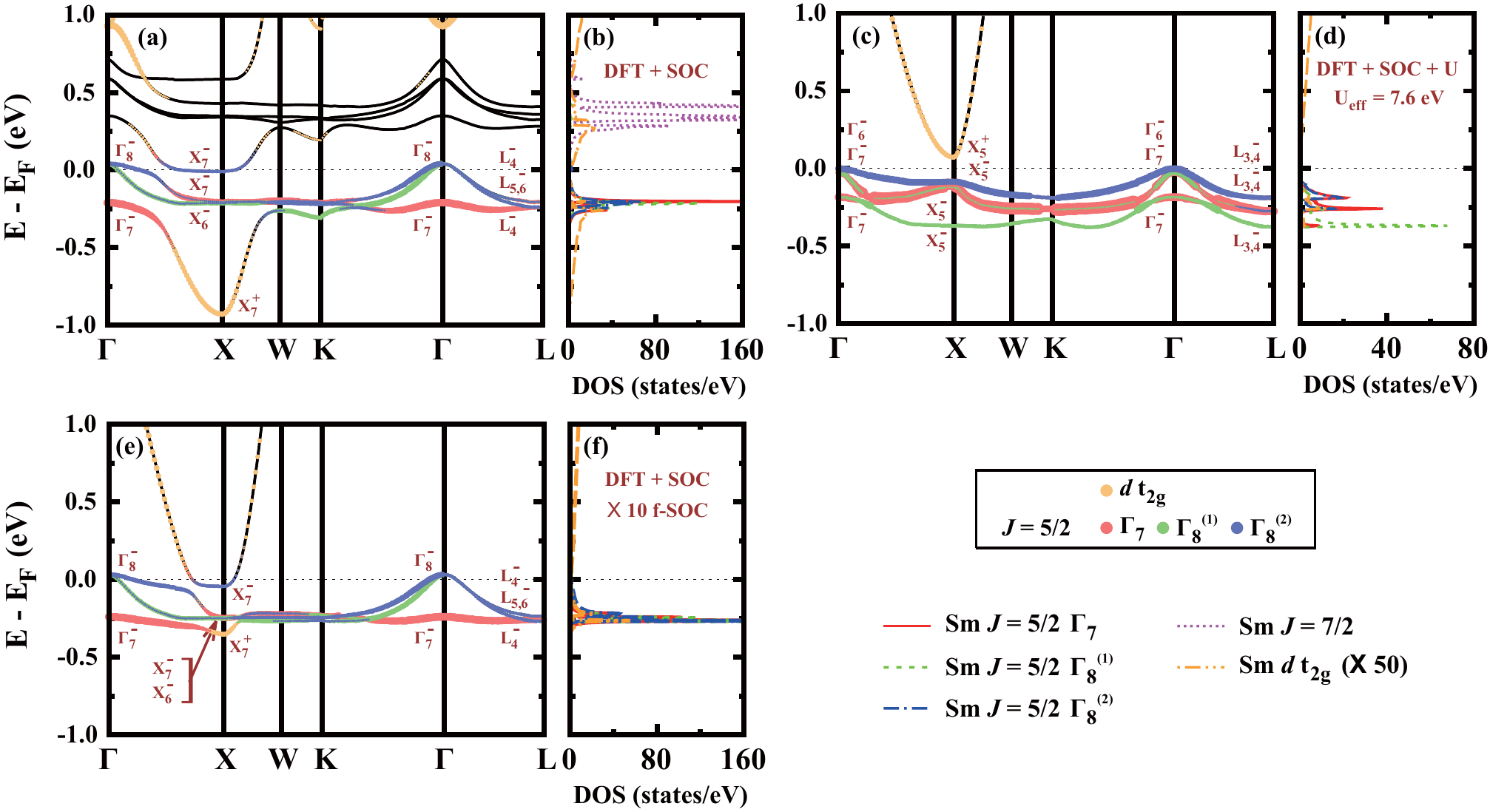}
\caption{(Color Online)
Band structures and DOSs of a black phase of SmS ($a$ = 5.953 $\AA$) in the DFT schemes.
(a) Symmetry decomposed band structure with the relativistic double group bases
in the GGA + SOC scheme, which gives erroneously the slightly overlapped semimetallic state.
Sm 5$d$ $t_{2g}$ band is seen to be hybridized with $\Gamma_{7}$ doublet bands.
The valence state of Sm is obtained to be 2.56+ in this case.
(b) Symmetry projected partial DOS (PDOS) in the GGA + SOC scheme.
Sm $J = 5/2$ DOS are projected into the double group bases.
Sm $5d$ $t_{2g}$ DOS is magnified 50 times for a clear view.
(c) and (d) present symmetry decomposed band structure and PDOS
in the GGA + SOC + $U$ ($U_{eff}$ = 7.6 eV) scheme,
which seems to describe correctly a semiconducting b-SmS with the band gap of 87 meV.
The valence state of Sm is obtained to be 2.23+.
However, band symmetries at high symmetry {\bf $k$}-points in this scheme are different
from those in the GGA + SOC scheme.
(e) and (f) show symmetry decomposed band structure and PDOS
in the GGA + SOC with 10 times enhanced SOC of Sm $4f$-electron (10$\times f$-SOC scheme).
Due to the large SOC strength, the $J = 7/2$ band is shifted up to a higher energy side.
In contrast to the GGA + SOC + $U$ scheme, band symmetries at high symmetry {\bf $k$}-points in this scheme
preserve those of the GGA + SOC scheme.
Sm $t_{2g}$ band hybridizes with $\Gamma_{7}$ doublet, resulting in the hybridization gap near $X$.
The valence state of Sm is obtained to be 2.27+.
}
\label{black-DFT}
\end{figure*}

Figure~\ref{black-DFT} shows the band structures and densities of states (DOSs)
of a black phase of SmS (b-SmS) obtained by the DFT schemes.
Figure~\ref{black-DFT}(a) shows symmetry decomposed band structure of b-SmS
in the GGA + SOC scheme (SOC: spin-orbit coupling).
Since Sm 4$f$-electrons feel much larger SOC than the cubic crystal field,
the SOC bases incorporating the cubic crystal field should be
utilized~\cite{Kang13,Pappalardo61}.
The cubic crystal field splits $4f_{5/2}$ states at $\Gamma$ into
lower $\Gamma_{7}$ doublet and higher $\Gamma_{8}$ quartet,
as in SmB$_{6}$~\cite{Kang13}.
Sm 4$f$ bands are dominant near the Fermi level ($E_F$), and
$4f_{5/2}$ and $4f_{7/2}$ states are located mainly
below and above $E_F$, respectively.
The splitting between $4f_{5/2}$ and $4f_{7/2}$ states is about 0.75 eV,
which is related to the strength of the SOC of 4$f$ electrons in Sm ion.
In fact, most of $4f_{7/2}$ states are to be shifted up in energy
due to the strong correlation effect of 4$f$-electrons.

Sm $5d$ $t_{2g}$ characters are also shown in order to visualize
the hybridization between the $\Gamma_{7}$ doublet and $t_{2g}$ band.
Since the lobes of $e_{g}$ and $t_{2g}$ orbitals of Sm $5d$ electrons
are along and away from the anion sulfur ions, respectively,
the energy of $t_{2g}$ band is to be lower than that of $e_{g}$ band.
This feature in SmS is contrary to that in SmB$_{6}$, for which
the $e_{g}$ band is lower in energy than the $t_{2g}$ band
because the lobes of $t_{2g}$ orbitals are toward anion boron ions~\cite{Kang13}.
Therefore, in SmB$_{6}$, the $f$-$d$ hybridization occurs
between the $\Gamma_{7}$ doublet and $e_{g}$ band,
while, in SmS, it occurs between the $\Gamma_{7}$ doublet and $t_{2g}$ band.
At $X$, the $t_{2g}$ band is located below $4f_{5/2}$ bands,
and so the order of parities is changed.
The valence state of Sm in the GGA + SOC scheme is estimated to be 2.56+,
which is far from the semiconducting black phase of the valence state of Sm, 2+.
Furthermore, the band gap is not obtained which is contrary to the experiment.

Figure~\ref{black-DFT}(b) presents the symmetry-projected partial DOSs (PDOSs)
of b-SmS in the GGA + SOC scheme.
Sm $4f_{5/2}$ DOS is projected into the relativistic double group bases
and Sm $4f_{7/2}$ and $5d$ $t_{2g}$ PDOSs are also provided.
It is seen that the DOS at $E_F$ is not zero, implying that the system is not
an insulator. Rather it is semimetallic, as shown in Fig.~\ref{black-DFT}(a).

Figure \ref{black-DFT}(c) and (d) show electronic structures of b-SmS
in the GGA + SOC + $U$ scheme ($U$: Coulomb correlation of Sm $4f$ electrons).
When $U_{eff} = 7.6$ eV, it gives an insulating phase with the band gap of 87 meV,
which is quite comparable to the experimental band gap of $\sim$0.15 eV~\cite{Varma76}.
Once the semiconducting phase of SmS is obtained,
the parity inversion does not happen, which results
in the trivial $\mathbb{Z}_{2}$ topological number.
The valence state of Sm is estimated to be 2.23+, which is close to 2+.
However, the band symmetries at high symmetry {\bf $k$}-points are
different from those in the GGA + SOC scheme,
which turns out to be wrong
in view of the DMFT results (see Fig. 2(c) in the main text).

Figure~\ref{black-DFT} (e) and (f) show symmetry decomposed band structure and PDOS of b-SmS
in the GGA + SOC with 10 times enhanced SOC of Sm $4f$-electron (10$\times f$-SOC scheme).
Due to the artificially large SOC strength, the $4f_{7/2}$ bands are shifted up to the higher energy side.
In contrast to the GGA + SOC + $U$ scheme, the band symmetries at high symmetry {\bf $k$}-point
in the 10$\times f$-SOC scheme are consistent with those of the GGA + SOC scheme.
Sm $5d$ $t_{2g}$ band hybridizes with $\Gamma_{7}$ doublet, resulting in the hybridization gap near $X$.
The valence state of Sm is estimated to be 2.27+.

Thus the DFT schemes can not describe the ground state electronic structure of b-SmS properly.
Also, according to the angle-resolved photoemission spectroscopy (ARPES) experiment~\cite{Ito02},
Sm $4f$ bands are very flat and distributed broadly from 1 to 4 eV below $E_F$,
which is different from the DFT results.
This is due to the strong correlation effect in $4f$ electrons,
which cannot be captured in conventional DFT calculations.

\subsection{DMFT band structure}

\begin{figure}[t]
\includegraphics[width=8.5 cm]{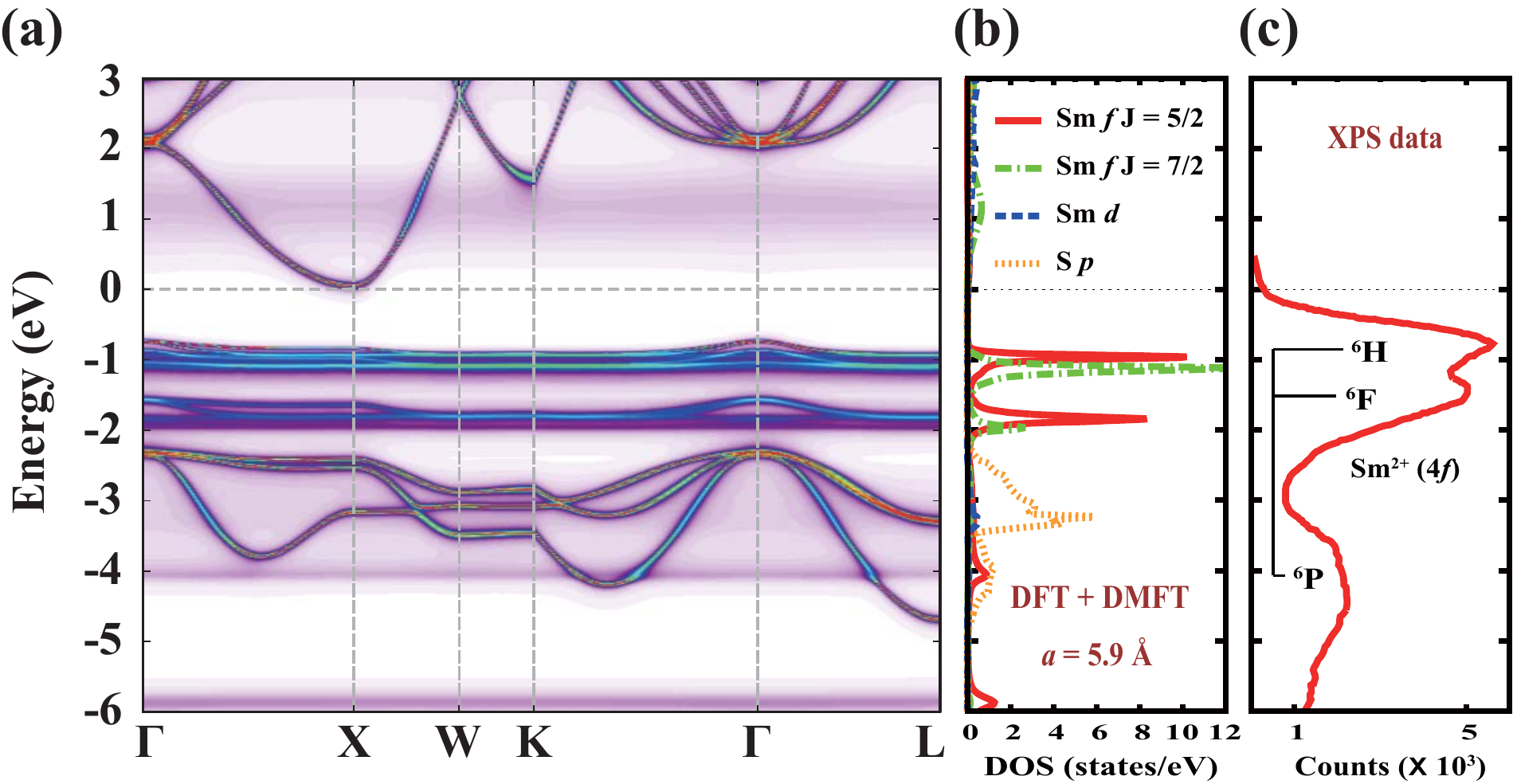}
\caption{(Color Online) Electronic structure of b-SmS.
(a) The DMFT band structure of insulating b-SmS ($a$ = 5.9 $\AA$)
at $T = 11.6$ K.
On-site Coulomb and exchange energies of $U = 6.1$ eV and $J = 0.8355$ eV were adopted.
(b) Partial DOS of b-SmS.
(c) XPS data for b-SmS taken from Ref.~\cite{Campagna74}.
}
\label{black-DMFT}
\end{figure}

Figure~\ref{black-DMFT} shows the band structure and DOS of b-SmS ($a$ = 5.9 $\AA$)
obtained by the charge self-consistent DFT + DMFT scheme
at $T = 11.6$ K.
It is seen that the insulating electronic structure is correctly
described by the DMFT scheme.
The obtained band gap and the valence configuration of Sm are 0.46 eV and 2.06+,
respectively.
The band structure in Fig.~\ref{black-DMFT} is very similar to
that obtained by the periodic Anderson model calculation \cite{Lehner98},
and consistent with ARPES data \cite{Ito02}.
The partial DOS in Fig.~\ref{black-DMFT}b shows 4$f$ multiplets,
which are in good agreement with XPS data \cite{Campagna74}.
Since the Sm $d$-band is above the Sm $f$-band,
b-SmS is a topologically trivial insulator.

\section{A golden phase of SmS (g-SmS)}
\subsection{Band structure in the DFT + SOC with 10$\times f$-SOC scheme}

\begin{figure}[t]
\includegraphics[width=8.5 cm]{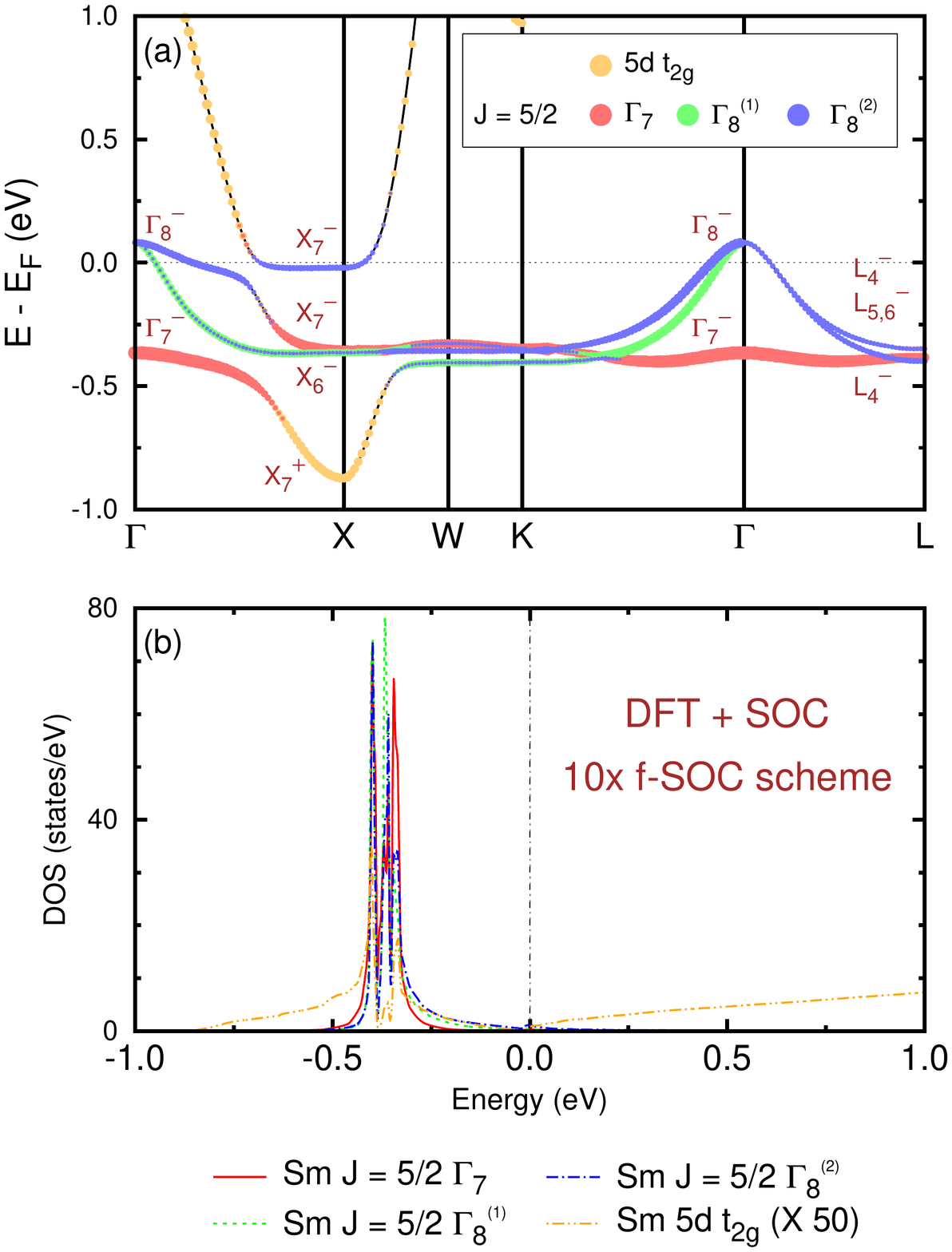}
\caption{(Color Online)
Band structure and PDOS of golden phase of SmS (g-SmS) ($a$ = 5.6 $\AA$)
in the GGA + SOC with 10$\times f$-SOC scheme.
(a) Symmetry decomposed band structure with the relativistic double group bases.
Sm 5$d$ $t_{2g}$ band is seen to be hybridized with $\Gamma_{7}$ doublet bands.
The valence state of Sm is obtained to be 2.36+ in this case.
(b) Symmetry projected PDOS.
Sm $J = 5/2$ DOS is projected into the double group bases.
Sm $5d$ $t_{2g}$ DOS is magnified 50 times for a clear view.
}
\label{golden-DFT}
\end{figure}

Figure~\ref{golden-DFT} provides electronic structures of g-SmS ($a$ = 5.6 $\AA$)
in the GGA + SOC with 10$\times f$-SOC scheme.
The electronic structures in Fig.~\ref{golden-DFT} are quite similar
to those of g-SmS described by the DMFT at low $T$ (see Fig. 2(c) in the main text).
But, the number of $f$-electron ($n_{f}$) in the DFT + SOC with 10$\times f$-SOC scheme
is 5.64 (Sm$^{2.36+}$), while $n_{f}$ in the DMFT scheme is 5.27 (Sm$^{2.73+}$).

\subsection{Temperature-dependent DMFT physical parameters}

\begin{figure*}[t]
\includegraphics[width=18 cm]{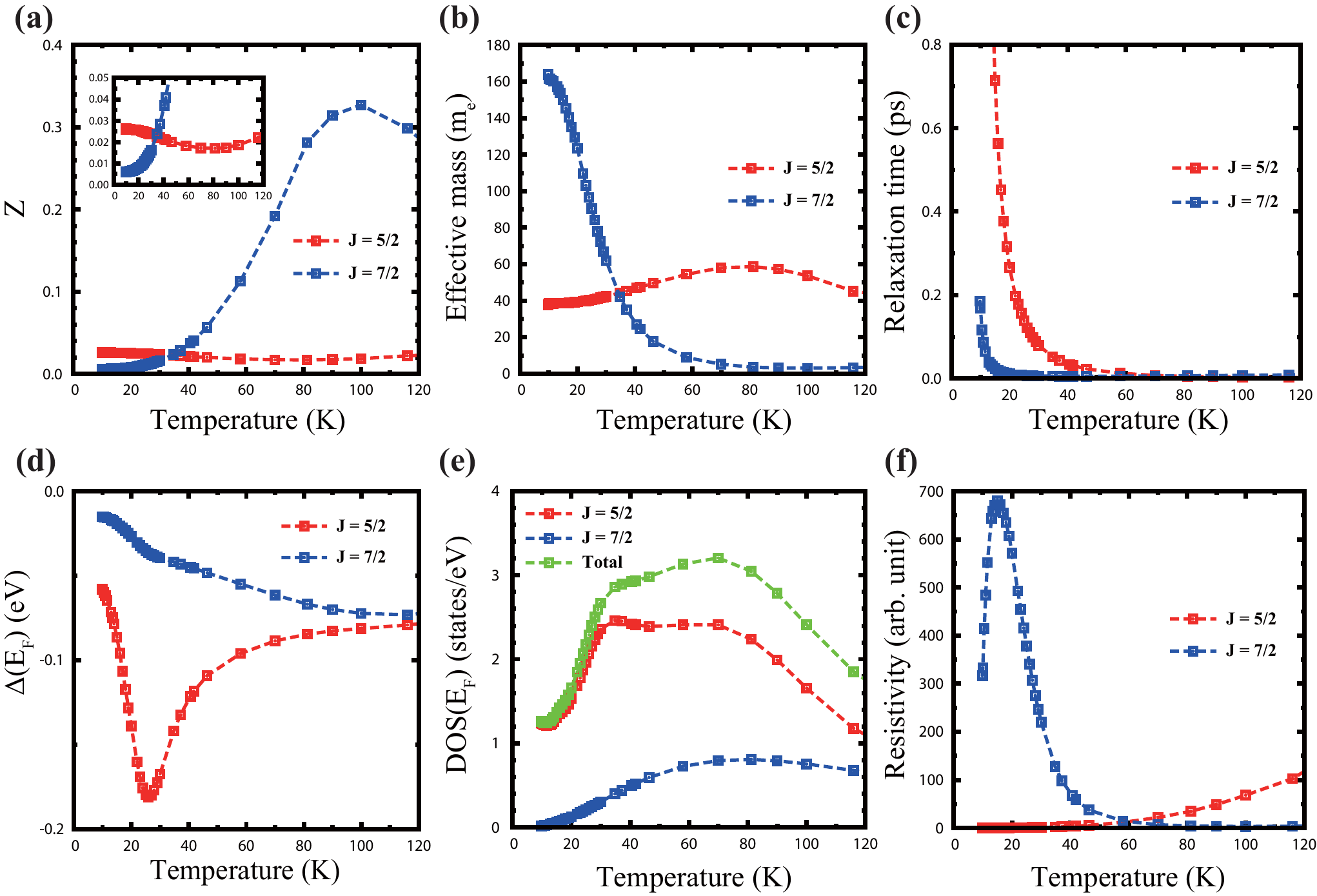}
\caption{(Color Online)
$T$-dependent DMFT physical parameters in g-SmS ($a$ = 5.6 $\AA$).
(a) and (b) represent the renormalization factor $Z$ and effective mass $m^*$ versus $T$.
Upon cooling, $Z$ of $J = 5/2$ increases, while that of $J = 7/2$ decreases.
(Inset shows this trend clearly.)
The behavior of $m^*$ is just opposite, because $m^*=1/Z$.
(c) Relaxation time versus $T$.
Both relaxation times of $J = 5/2$ and $J = 7/2$ diverge at low $T$,
suggesting the crossover to the Landau Fermi liquid behavior.
(d) $T$-dependent behavior of the hybridization function at $E_{F}$ ($\Delta(E_{F})$).
It has the largest value near $T = 25$ K to produce the strong Kondo screening behavior.
(e) DOS at $E_{F}$ versus $T$.
The total DOS at $E_{F}$ increases first and then decreases below $T \simeq 70$ K.
(f) Resistivity versus $T$ obtained by the Drude model.
The resistivity drop for $J = 7/2$ is clearly shown at $T \simeq 13$ K,
which is also captured in the DMFT result (see Fig. 3(c) in the main text).
}
\label{T-dep}
\end{figure*}

Figure \ref{T-dep} presents the temperature ($T$)-dependent physical properties
of g-SmS ($a$ = 5.6 $\AA$) described by the DMFT.
Figure~\ref{T-dep}(a) and (b) show the renormalization factor $Z$
and effective mass $m^{*}$ versus $T$.
Upon cooling, $Z$ of $J = 5/2$ ($J = 7/2$) increases (decreases),
resulting in decreasing (increasing) $m^{*}$ of $J = 5/2$ ($J = 7/2$).
Relaxation times of both $J = 5/2$ and $J = 7/2$ diverge upon cooling,
as shown in Fig.~\ref{T-dep}(c),
suggesting the crossover to the Landau Fermi liquid behavior at low $T$.

Figure~\ref{T-dep}(d) provides the $T$-dependent behavior of
the hybridization function at $E_{F}$ ($\Delta(E_{F})$), which is closely related
to the Kondo interaction strength. Hence the largest $\Delta(E_{F})$ value near $T \simeq 25$ K
suggests that Kondo screening behavior becomes strongly  enhanced at this temperature range.
On the other hand, the total DOS at $E_{F}$ in Fig.~\ref{T-dep}(e) increases first
and then decreases below $T \simeq 70$ K.
Note that the total DOS at $E_{F}$ is saturated at low-$T$,
indicating the full lattice coherence below $T \simeq 13$ K.

Fig.~\ref{T-dep}(f) shows the resistivity versus $T$ obtained by the Drude model.
The resistivity drop for $J = 7/2$ is clearly shown below $T \simeq 13$ K,
which is also captured in the DMFT result (see Fig. 3(c) in the main text).
It implies that the crossover from the Kondo liquid to the Landau Fermi liquid
occurs below $T \simeq 13$ K.

The images for moving pictures
containing $T$-dependent band structures and Fermi surfaces
(T-band-movie1 and T-FS-movie2) are provided
for the additional Supplementary Information.

\end{document}